\newif\ifTwoColumn
\newif\ifSUBMIT
\newif\ifCOMMENTS
\newif\ifFIGs
\newif\ifFIGoneColumn
\let\ifSUBMIT\iftrue
\let\ifCOMMENTS\iffalse
\let\ifFIGoneColumn\iftrue
\documentclass[aip,reprint,jcp,amsmath,amssymb,superscriptaddress,floatfix]{revtex4-1}

\usepackage{graphicx}
\usepackage{amsmath}
\usepackage{xcolor}
\usepackage{hyperref}
\usepackage{array,mathtools,amssymb,booktabs}
\newcolumntype{C}{>{$}c<{$}}
\AtBeginDocument{
\heavyrulewidth=.10em
\lightrulewidth=.05em
\cmidrulewidth=.03em
\belowrulesep=.65ex
\belowbottomsep=0pt
\aboverulesep=.4ex
\abovetopsep=0pt
\cmidrulesep=\doublerulesep
\cmidrulekern=0.5em
\defaultaddspace=.5em
}

\usepackage{array,mathtools,amssymb,booktabs}

\ifSUBMIT
  \ifCOMMENTS
    \usepackage{color}    %%% NOTE this is necessary in pdflatex
    \usepackage[normalem]{ulem}
    \def\EDITS#1{{\color{green}#1}}
    \def\STRIKE#1{{\color{red}\sout{#1}}}
    
    \def\NSTRIKE#1{{\color{blue}\sout{#1}}}
  \else
    \def\EDITS#1{#1}
    \def\STRIKE#1{}
    
    \def\NSTRIKE#1{}
  \fi
\else
  \usepackage{color}    %%% NOTE this is necessary in pdflatex
  \usepackage[normalem]{ulem}
 \definecolor{mygreen}{RGB}{0,180,0}    %%% NOTE that this isn't necessary.
  \def\EDITS#1{{\color{mygreen}#1}}
  \def\STRIKE#1{{\color{red}\sout{#1}}}

  \def\NSTRIKE#1{{\color{blue}\sout{#1}}}
\fi

\definecolor{mygray}{RGB}{128,128,128}

\DeclareMathOperator{\argmin}{arg\,min}

\usepackage{acronym}

\begin{document}
%%%%%%%%%%%%%%%%%%%%%%%%%%%%%%%%%%%%%%%%%%%%%%%%%%%%%%%%%%
\newlength\figurewide
\ifFIGoneColumn
  \figurewide=.5\columnwidth
\else
  \figurewide=.9\columnwidth
\fi

\title{Chemical dynamics 
between wells
across a time-dependent barrier:
Self-similarity in the Lagrangian descriptor
and reactive basins}
\author{Andrej Junginger}
\author{Lennart Duvenbeck}
\author{Matthias Feldmaier}
\author{J\"org Main}
\author{G\"unter Wunner}
\affiliation{%
Institut f\"ur Theoretische Physik 1, 
Universit\"at Stuttgart, 
70550 Stuttgart,
Germany}

\author{Rigoberto Hernandez}
\email[Correspondence to: Rigoberto Hernandez, Department of Chemistry,
Johns Hopkins University,
Baltimore, MD 21218.  E-mail: ]{r.hernandez@jhu.edu}
\affiliation{%
Department of Chemistry,
Johns Hopkins University,
Baltimore, \mbox{MD 21218, USA}}%}

\date{\today}
%%% user=defined commands %%%%%%%%%%%%%%%%%%%%%%%%%%%%%%%%%%%%%%%%%%%%%%%%%%%% 
\newcommand{\EQ}{Eq.}
\newcommand{\EQS}{Eqs.}
\newcommand{\FIG}{Fig.}
\newcommand{\FIGS}{Figs.}
\newcommand{\REF}{Ref.}
\newcommand{\REFS}{Refs.}
\newcommand{\SEC}{Sec.}
\newcommand{\SECS}{Secs.}
\newcommand{\eg}{e.\,g.}
\newcommand{\cf}{cf.}
\newcommand{\ie}{i.\,e.}
\newcommand{\ud}{\mathrm{d}}
\newcommand{\ue}{\mathrm{e}}
\newcommand{\kB}{k_\mathrm{B}}
\newcommand{\VLiCN}{V_\mathrm{LiCN}}
\newcommand{\VCN}{V_\mathrm{C-N}}
\newcommand{\VLi}{V_\mathrm{Li-CN}}
\renewcommand{\vec}[1]{\boldsymbol{#1}}
\newcommand{\qq}{\vec{q}}
\newcommand{\xx}{\vec{x}}
\newcommand{\vv}{\vec{v}}
\newcommand{\transpose}{\mathsf{T}}
\newcommand{\reactantpop}{\mathcal{P}}
\newcommand{\kf}{k_\mathrm{f}}
\newcommand{\etal}{\emph{et al.}}
\newcommand{\LD}{\mathcal{L}}
\newcommand{\LDbo}{\mathcal{L}_\text{bo}}
\newcommand{\LDwo}{\mathcal{L}_\text{wo}}
\newcommand{\LDf}{\LD^\text{(f)}}
\newcommand{\LDb}{\LD^\text{(b)}}
\newcommand{\LDfb}{\LD^\text{(fb)}}
\newcommand{\LDfbw}{\LD^\text{(fbw)}}
\newcommand{\Ws}{\mathcal{W}_\text{s}}
\newcommand{\Wu}{\mathcal{W}_\text{u}}
\newcommand{\Wsu}{\mathcal{W}_\text{s,u}}
\newcommand{\TSt}{\mathcal{T}}
\newcommand{\weightingf}{\chi^\text{(f)}}
\newcommand{\weightingb}{\chi^\text{(b)}}
\newcommand{\weightingfb}{\chi^\text{(f,b)}}
\newcommand{\vtherm}{v_\text{therm}}
\newcommand{\comment}[1]{\textsf{\textcolor{orange}{[#1]}}}
\newcommand{\sno}[1]{_\mathrm{#1}}
\newcommand{\no}[1]{\mathrm{#1}}
\newcommand{\acnew}[1]{\acfi{#1}\acused{#1}}
\newcommand{\VMorse}{V_\text{Morse}}
\newcommand{\VGauss}{V_\text{Gauss}}
\newcommand{\Eg}{E_\text{G}}
\newcommand{\xb}{x_\text{b}}
\newcommand{\subbo}{_\text{bo}}
\newcommand{\subwo}{_\text{wo}}
\newcommand{\xtsr}{\bar{x}}

\begin{abstract}\label{sec:abstract}
In chemical or physical reaction dynamics, it is essential
to distinguish precisely between reactants and products for all time.
This task is especially demanding in time-dependent 
or driven systems because therein the 
dividing surface (DS) between these states often exhibits a 
nontrivial time-dependence.
The so-called transition state (TS) trajectory 
has been seen to define a DS which
is free of recrossings in a large number of
one-dimensional reactions across time-dependent barriers,
and, thus, allows one to determine exact reaction rates.
A fundamental challenge to applying this method is the construction of the TS 
trajectory itself.
The minimization of Lagrangian descriptors (LDs) 
provides a general and powerful scheme to obtain 
that trajectory even when perturbation theory fails.
Both approaches encounter possible breakdowns when the overall potential is
\EDITS{bounded}, admitting the possibility of returns to the barrier long
after trajectories have reached the product or reactant wells.
Such global dynamics cannot be captured by perturbation theory.
Meanwhile, in the LD-DS approach, 
it leads to the emergence of 
additional local minima which make it difficult to extract 
the optimal branch associated with the desired TS trajectory.
In this work, we illustrate this behavior for a time-dependent 
double-well potential revealing a 
self-similar structure of the LD, and we demonstrate how the 
reflections and side-minima can be
addressed by an appropriate modification of the LD
\EDITS{associated with the direct rate across the barrier.}
\end{abstract}
\maketitle

%%%%%%%%%%%%%%%%%%%%%%%%%%%%%%%%%%%%%%%%%%%%%%%%%%%%%%%%%%%%%%%%%%%%%%%%%%%%%%%%
\section{Introduction}

One of the grand challenges in the field of driven reaction dynamics 
is the complete characterization of the rates and pathways so as
to allow for control.
One possible route for such driving is the perturbation of the underlying 
potential energy surface by time-dependent, external fields.
\cite{Yamanouchi2002,hern05a,Sussman2006,Kawai07,Kawai11laser,Keshavamurthy2009,
Keshavamurthy2015,Revuelta2015}
The configurational change of the reactive system is typically 
mediated by an energy barrier 
separating reactant and product basins which must be somehow
affected by the external control mechanism.
In the limit of no driving,
transition state theory (TST)
\cite{pitzer,pechukas1981,truh79,truh85,truhlar91,truh96,truh2000,
Komatsuzaki2001,Waalkens2008,hern08d,Komatsuzaki2010,hern10a,Henkelman2016} 
provides a powerful, though usually approximate, 
framework to calculate the rate
from the reactive 
flux though the dividing surface (DS) separating the reactant and 
product regions.
Such rates are exact if the DS is crossed by each 
reactive trajectory exactly once.
Thus a central task for applying TST is 
the determination of a DS with this no-recrossing property.
In time-independent systems with a two-dimensional configuration space, the DS 
is associated with an unstable periodic orbit at the barrier top, and in 
higher-dimensional systems it can be constructed using a normally hyperbolic 
invariant manifold.
\cite{pollak78,pech79a,hern93b,hern94,Jaffe00,Koon00,Jaffe02,Uzer02,Waalkens04b,Jaffe05,
Li06prl,Teramoto11,Waalkens13}
By contrast, in time-dependent systems, the DS is, in general, also 
time-dependent and the transition state (TS) trajectory,
\cite{dawn05b,dawn05a,hern06d,hern14b,hern14f,hern15a,Kawai2009a} 
which is a hyperbolic trajectory close to the barrier top, 
has proven to give rise to an associated 
non-recrossing time-dependent DS.

The TS trajectory can be constructed through perturbation theory
in several limiting cases,\cite{hern93b,dawn05a}
but the approach does not have an obvious zeroth-order reference in
barrier reactions\cite{hern16a}
and can break down when the dynamics is affected by features on the
potential energy surface far from the barrier region.
It has recently been shown that the minimization of Lagrangian descriptors 
\cite{Mancho2010,Mancho2013} 
(LDs) provides a general and powerful construction scheme to obtain the TS 
trajectory in such difficult cases.\cite{hern15a,hern15e,hern16a,hern16d} 
In simple terms, the initial condition for the TS trajectory
is the one for which the LD,
integrated for some sufficiently long time,
is a minimum over the
domain of the underlying phase space coordinates. 

The present work revisits
the prototypical reaction 
in a time-dependent double-well 
potential with oscillating barrier position
from our earlier work.\cite{hern16h} 
In this paper, we address 
a possible challenge to the minimization 
in the LD-DS procedure 
arising from the reflections of particles 
when the overall potential has characteristic bound 
reactant and product wells.
Without such wells, particles leaving sufficiently
far from the reaction region escape into a deep exit
channel from which they never return.
When the overall potential is 
\EDITS{bounded}, however, 
the global motion gives rise to (uncorrelated)
returns back to the 
reaction region despite the fact that the particles
relaxed into a given basin in the intervening time.
This effect leads, in general, 
to several or formally an infinite number 
of local LD minima, 
making it difficult to identify the 
optimal LD minima associated with the TS trajectory.
We demonstrate in this paper that the LD surface in phase space exhibits a 
huge number of local minima which are related to trajectories leaving and 
reentering the barrier region.
Equally important, we also show how to identify the
primary minimum needed to construct the TS trajectory 
\EDITS{for calculating direct rates},
and thereby resolve the use of the LD-DS method
for typical \EDITS{(bounded)} chemical reactions.

The extended LD-DS method needed to account for global recrossings
is developed in \SEC~\ref{sec:theory}.
Specifically, the LD is modified so that only a single 
minimum remains,
and it is precisely the one corresponding to 
the TS trajectory.
\EDITS{It corresponds to the imposition of reactant and product surfaces
defining the entry into the corresponding regions in
the reactive flux time correlation function for the rate, and which
are imposed to avoid recurrences from the reverse reaction.
Such an approach has been used routinely in the chemistry community,
and is beautifully imposed in the recent work of
Mesele and Thompson\cite{thompson16}
in obtaining expressions for the activation energy.}
This section also includes
the structure and parameters of the model double-well used 
in this work to illustrate the LD-DS method.
In \SEC~\ref{sec:results}, we investigate the self-similar LD 
structure in detail with respect to the model system.
The trajectories associated with the local minima are identified
clearly using the modified LD, and the 
optimal one 
leads to a numerical construction of 
the TS trajectory.
This central result is essential to the 
extension of the LD-DS method towards 
higher-dimensional, finite reactive systems.
The latter necessarily includes global recrossings which
would bedevil perturbation theory and the naive LD-DS approach.

\section{Theory}\label{sec:theory}

%%%%%%%%%%%%%%%%%%%%%%%%%%%%%%%%%%%%%%%%%%%%%%%%%%%%%%%%%%%%%%%%%%%%%%%%%%%%%%%
\begin{figure}[t]
\includegraphics[width=\columnwidth]{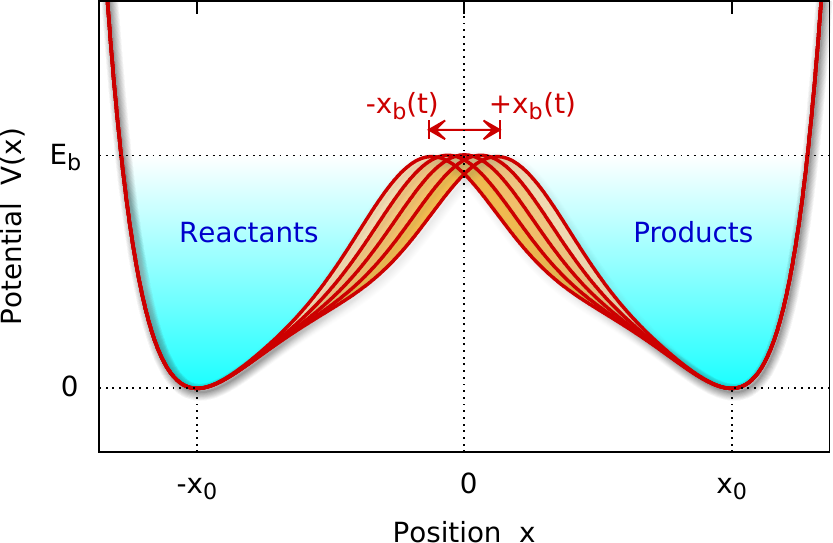}
\caption{%
Visualization of the time-dependent double-well potential in 
\EQS~\eqref{eq:potential}--\eqref{eq:potential-single-terms} at
different positions $\xb(t)=0,\pm0.2,\pm0.4$ of the barrier top.
}
\label{fig:potential}
\end{figure}
%%%%%%%%%%%%%%%%%%%%%%%%%%%%%%%%%%%%%%%%%%%%%%%%%%%%%%%%%%%%%%%%%%%%%%%%%%%%%%%

%%%%%%%%%%%%%%%%%%%%%%%%%%%%%%%%%%%%%%%%%%%%%%%%%%%%%%%%%%%%%%%%%%%%%%%%%%%%%%%
\begin{figure*}[t]
\centering
\includegraphics[width=\textwidth]{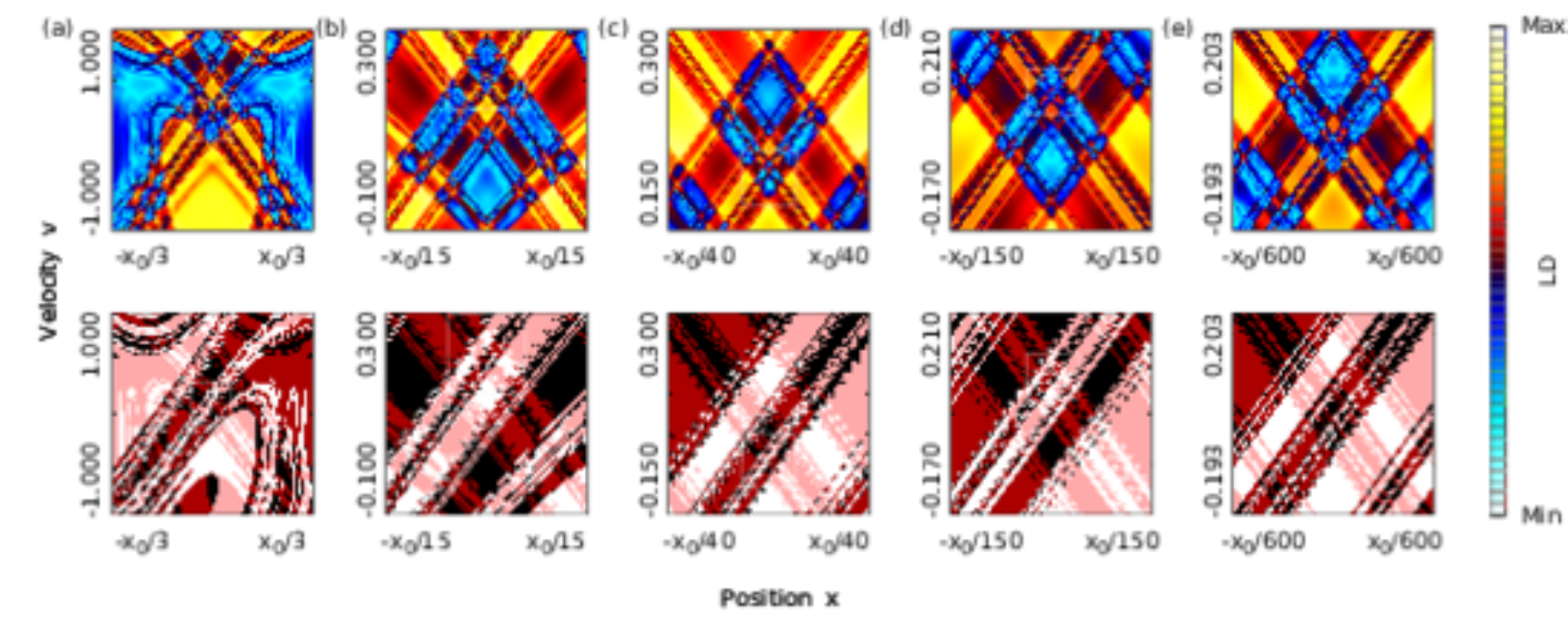}
\caption{%
Phase space portraits of the LD~\eqref{eq:LD} (top row) as well as reactive 
basin portraits (bottom row) in the barrier region for a fixed integration time 
$\tau=20$ covering ten oscillation periods of the barrier top.
From panels (a) to (e), the phase space region is magnified and the zoomed area 
is indicated by the white rectangles.
[The range of the $x$-axis is given relative to the well position $x_0$ which 
corresponds to absolute values of (a) $\pm1.0$, (b) $\pm0.2$, (c) $\pm0.75$, 
(d) $\pm0.02$, and (e) $\pm0.005$.
As an orientation for the  magnitude of the velocity, we note that typical 
trajectories crossing the saddle reach velocities up to $v\approx2.0$ at the 
bottom of the well while the maximum velocity of the TS trajectory is 
$v\approx0.02$.]
The different zoom levels show recurring structures indicating a self-similar 
LD structure close to the TS.
In the bottom row, the reactive phase space regions are visualized: 
Black and white regions indicate nonreactive particles which either start and 
end in the left (black) or right (white) basin.
By contrast, the red regions show reactive particles which undergo forward 
(dark red) or backward (light red) reactions.
Each of the plots shows the respective phase space region with a resolution 
of $1000\times1000$ pixels.
}
\label{fig:LD-basin}
\end{figure*}
%%%%%%%%%%%%%%%%%%%%%%%%%%%%%%%%%%%%%%%%%%%%%%%%%%%%%%%%%%%%%%%%%%%%%%%%%%%%%%%

In this paper, we investigate the time-dependent double-well potential 
(see \FIG~\ref{fig:potential})
\begin{equation}
  V(x,t) = 
  \VMorse^-(x) + \VGauss(x,t) + \VMorse^+(x) 
  \label{eq:potential}
\end{equation}
that has already been subject of the work in \REF~\onlinecite{hern16h}.
This potential consists of two Morse potentials $\VMorse^\pm$ providing 
the reactant ($-$) and product ($+$) wells and a time-dependent Gaussian 
potential $\VGauss$ serving as the time-dependent barrier.
Specifically, these dimensionless potential terms are
\begin{subequations}
\begin{align}
\VMorse^\pm(x) &= D \left[1 - \ue^{ \pm b (x \mp x_0)}\right]^2  \,, \\
\VGauss (x,t) &= D \, \ue^{- a \left[ x - x_{\rm b}(t) \right]^2} \,,
\end{align}%
\label{eq:potential-single-terms}%
\end{subequations}
where $\xb$ is the position of the barrier, 
the factor $D$ determines the energy scale, 
and $\pm x_0$ are the positions of the wells.
As in \REF~\onlinecite{hern16h}, we use $D=1$ for the energy scale, and 
specify the spacial scale of the system by setting 
$x_0=3$ as well as $a=b=1$.
Moreover, we apply a sinusoidal driving of the barrier 
according to $\xb(t)= 0.4 \sin( \pi t )$, so that the barrier oscillation is 
$0.4/3\approx13\%$ of the distance to the well's minimum.
For a centered barrier top at $\xb(t)=0$, the barrier height is
\begin{equation}
E^\ddagger_{\xb(t)=0} = D (3 + 2 \ue^{-2 b x_0} - 4 \ue^{-b x_0})
\approx 2.806 \,,
\end{equation}
which varies only slightly during the oscillation of the barrier top.
For simplicity, we use mass-weighted coordinates throughout this paper.
Together with the length scale given by the position $x_0$ of the well minima 
and the barrier height $E^\ddagger$ as a reference for the energy scale, this 
defines the dimensionless units of the system.

As shown in 
\REFS~\onlinecite{hern15e,hern16a,hern16d}, 
a nonperturbative approach to constructing a time-dependent, recrossing-free DS 
is given by a minimization procedure of LDs with respect to the underlying phase 
space coordinates.
In the context of TST, the LD is defined by the integral
\begin{equation}
 \LD ( \vec x_0, \vec v_0, t_0) = 
 \int_{t_0-\tau}^{t_0+\tau} \| \vec v(t) \| \, \mathrm{d}t \,,
 \label{eq:LD}
\end{equation}
where $\vec v$ is the velocity of a certain trajectory and therefore 
$\LD$ is a measure of the trajectory's arc length over the time interval 
$[t_0-\tau;t_0+\tau]$.

The LD is of special importance for the reaction dynamics, because the stable 
and unstable manifolds $\Wsu$ attached to the (time-dependent) barrier top are 
directly related to the LD's 
forward  ($f$: $t_0\leq t \leq t_0+\tau$) and
backward ($b$: $t_0-\tau \leq t \leq t_0$) contribution, respectively, 
according to \cite{hern15e,hern16a,hern16d}
\begin{subequations}
\begin{align}
 \mathcal{W}_\text{s} (t_0) &= 
 \argmin \mathcal{L}^{\text{(f)}} (\vec x_0, \vec v_0, t_0) \,, \\
 \mathcal{W}_\text{u} (t_0) &= 
 \argmin \mathcal{L}^{\text{(b)}} (\vec x_0, \vec v_0, t_0) \,.
\end{align}%
\label{eq:Wsu}%
\end{subequations}
This is the case because the trajectories on these manifolds approach the 
barrier top in either forward or backward time yielding extremal properties of 
the LD.
In \EQ~\eqref{eq:Wsu}, the function `$\argmin$' denotes the value of the LD
argument, \ie~the respective phase space coordinates at the (local) minimum.

The coordinates of the TS trajectory $\TSt(t_0)$ at time $t_0$ are 
located on the intersection of these manifolds, 
\cite{hern15e,hern16a,hern16d} 
$\TSt(t_0) = \Ws(t_0) \cap \Wu(t_0)$,
so that it is directly related to the two-sided LD~\eqref{eq:LD} via
\begin{equation}
 \TSt(t_0) = \argmin \mathcal{L} (\vec x_0, \vec v_0, t_0) \,.
 \label{eq:LD-min}
\end{equation}
As we demonstrate below, the definition~\eqref{eq:LD} of the LD with 
\emph{fixed} integration time $\tau$ in a closed system with finite reactant and 
product wells has the disadvantage that the LD, in general, does not only 
exhibit a single minimum, but a huge number of minima.
This makes the identification and construction of the TS trajectory in 
\EQ~\eqref{eq:LD-min} difficult or even ambiguous.
The reason lies in the fact that when
a particle has an energy high enough to cross the 
barrier at least once, it will continuously be reflected to the barrier and 
undergo repeated barrier crossings 
(\EDITS{that is,} \emph{global} recrossings \EDITS{or recurrences}) 
\EDITS{which are not associated with the direct rate across the
barrier between reactant and product.}
\EDITS{Thus one needs to define the reaction region thorugh which
trajectories traverse from the reactant to product regions in order
to obtain the direct rae across the barrier.}
\EDITS{A} unique distinction between reactants and products can be made 
locally at the saddle (\EDITS{in terms of} \emph{local} crossings) by means of the TS trajectory.
To overcome the issue of global recurrences in the construction of the TS 
trajectory we 
modify the definition of the 
LD~\eqref{eq:LD} by simply replacing the fixed integration time $\tau$ 
with a variable one depending on the underlying trajectory.
For this purpose, we redefine the time interval over which 
the trajectory is integrated through the map:
\begin{equation}
  \Bigl[t_0-\tau,~t_0+\tau \Bigr] 
  ~\longrightarrow~
  \Bigl[ t_0-\tau^-[\xx(t)],~t_0+\tau^+[\xx(t)] \Bigr] \,.
  \label{eq:LD-time-interval-redef}
\end{equation}
Here, the values $\tau^\pm[\xx(t)]$ are trajectory-dependent integration times 
which we define according to
\begin{equation}
  \tau^\pm[\xx(t)] = \min \Bigl( \tau,~t \bigl.\bigr|_{|\xx(t)| > \xtsr} \Bigr)
  \label{eq:def-tau+-}
\end{equation}
with an appropriate ``size'' $\xtsr$ of the TS region. 
The redefinition of the integration time in \EQ~\eqref{eq:def-tau+-} 
is thereby limited by both $\tau$ and the time for the particle to
leave the barrier region (when $|x|$ is greater than $\xtsr$.)
\EDITS{The redefinition 
in \EQ~\eqref{eq:LD-time-interval-redef} may appear ad hoc, but is
motivated by the fact that all the undesired LD properties clearly result from 
recurrencs in the
trajectories being globally reflected to the barrier region. By construction,
our redefinition precisely retains only the
trajectories the contribute to the direct rate.}
The value of $\xtsr$ in \EQ~\eqref{eq:def-tau+-} needs to be appropriately defined:
On the one hand, it should not affect trajectories which do not leave the 
barrier; this suggests 
that it be a minimal value on the order of the barrier 
oscillation amplitude.
On the other hand, it should be large enough
to remove the effect of global recrossings
of particles returning from the reactant or product wells;
this suggests that its maximum value should be on 
the order of the distance between the barrier top and the well's minimum.
Finally, we note that the integration times $\tau^\pm$ may also differ in the 
forward and backward time direction.

\section{Results}\label{sec:results}

In this section, we apply the LD formalism, \EQ~\eqref{eq:LD-min}, to the 
double-well potential in \EQ~\eqref{eq:potential}.
In \FIG~\ref{fig:LD-basin}, we present the LD (top row) as well as the reactive 
basin portraits (bottom row) in phase space for a \emph{fixed} integration 
time $\tau=20$ and the panels \mbox{(a)--(e)} are different zoom levels close 
to the TS.
The reactive basin plots are 
coded through colors denoting
whether or not a particle is reactive:
black and white regions indicate nonreactive particles which either start and 
end in the reactant (black) or product (white) basin, and the red regions show 
reactive particles which undergo forward (dark red) or backward (light red) 
reactions.
The rich structure of the LD landscape, in the top left panel 
of \FIG~\ref{fig:LD-basin} for example, indicates the presence
of chaotic and regular regions in the dynamics.

The LD portraits show recurring 
structures at the different zoom levels \mbox{(a)--(e)} 
in \FIG~\ref{fig:LD-basin},
revealing a 
self-similar structure of the LD 
in a regular region of the phase space
(which we have further verified to be present down to the limit of numerical 
accuracy; not shown).
Equally important,
the reactive basin plots show the same 
recurring structure and the borders between the different reactive regions 
coincide with those observed in the LD plots.
This verifies previous results \cite{hern16d,hern16h} in which 
such a connection 
has been observed and extends them to the 
present case including reflections of the 
particles at the potential walls.

%%%%%%%%%%%%%%%%%%%%%%%%%%%%%%%%%%%%%%%%%%%%%%%%%%%%%%%%%%%%%%%%%%%%%%%%%%%%%%%
\begin{figure}[t]
\centering
\includegraphics[width=.8\columnwidth]{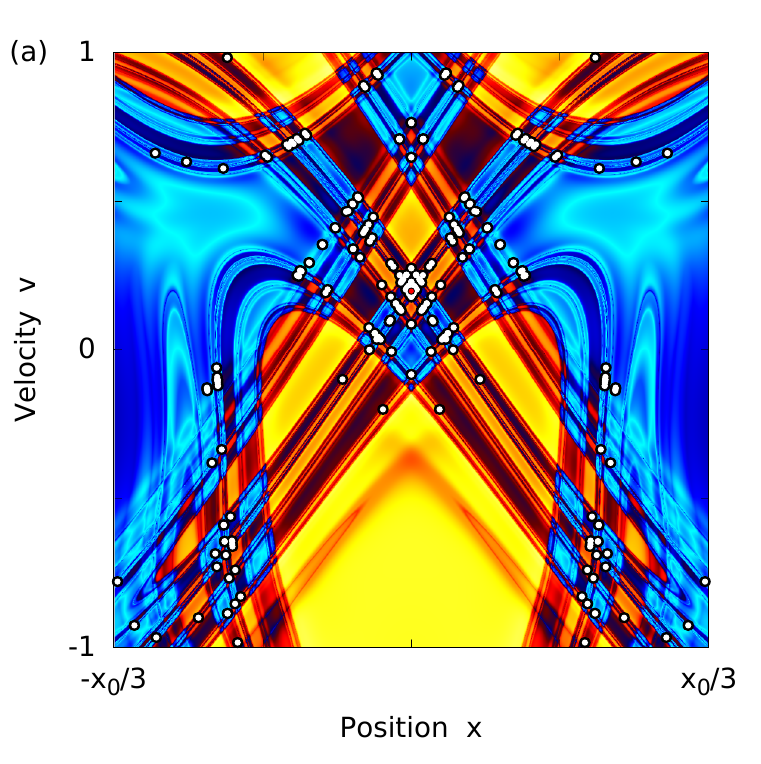}
\includegraphics[width=.85\columnwidth]{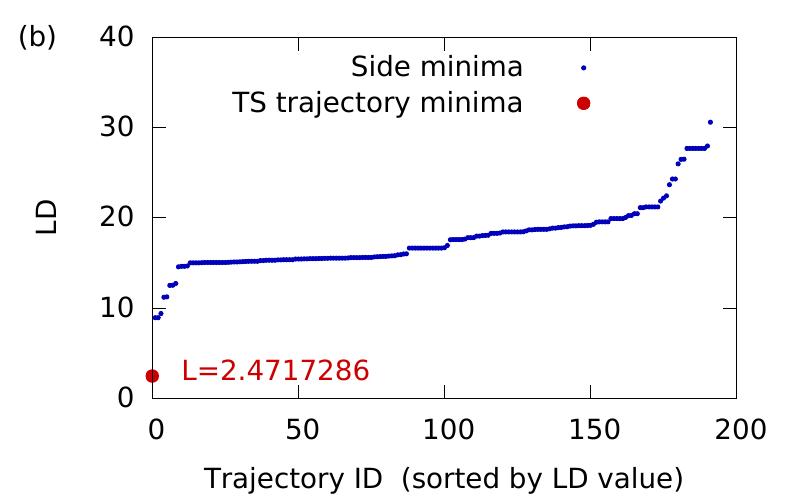}
\caption{%
(a)
LD phase space portrait for the system \eqref{eq:potential} with visualization 
of its (local) minima.
The selection of the latter is highlighted by the white dots and the minima are 
either observed at the LD minimum valleys [equivalent to the manifolds 
according to \EQ~\eqref{eq:Wsu}] or the intersections of these valleys.
(b)
Presentation of the LD value corresponding to the single local minima:
The minimum related to the TS trajectory (big red dot) has the smallest value 
and it is the one associated to the TS trajectory. 
}
\label{fig:minima}
\end{figure}
%%%%%%%%%%%%%%%%%%%%%%%%%%%%%%%%%%%%%%%%%%%%%%%%%%%%%%%%%%%%%%%%%%%%%%%%%%%%%%%

In \FIG~\ref{fig:minima}(a), we illustrate the LD portrait from 
\FIG~\ref{fig:LD-basin}(a) together with a selection of its (local) minima: 
Each of the white dots represents a local minimum of the LD as the result of 
a numerical search.
We note that the minima highlighted by the white dots are only a selection of
points.
As can be seen in the LD portrait, there are more intersections of the 
manifolds related to more local minima, but not all of them are highlighted in 
order to not overload the figure.
The minima are located either on one of the manifolds $\Wsu$ which are 
equivalent to LD minimum valleys according to \EQ~\eqref{eq:Wsu} or on their 
intersections, respectively.
However, even for this amount of local minima, it becomes obvious that 
\EQ~\eqref{eq:LD-min} is difficult to apply in order to locate the single 
minimum related to the TS trajectory.

Taking into account not only the relation~\eqref{eq:LD-min}, but also looking 
at the actual LD values of the minima can help to single out the desired 
minimum:
This can be understood from \FIG~\ref{fig:minima}(b) where the LD value of the 
local minima is visualized for the different trajectories.
(Note that, on the horizontal axis, the different trajectories are sorted 
ascending according to their LD value, so that this axis has no physical 
meaning).
The small blue dots show the LD value of all local minima and they exhibit a 
step structure with some steps indicating a big increase of the LD value.
From all the minima, there is one outstanding value (highlighted as a red dot 
with an LD value of $\LD=2.4717286$).
This value stands out because it is, by far, the smallest one with the 
second-smallest minimum exhibiting an LD value that is already three times as 
large.

%%%%%%%%%%%%%%%%%%%%%%%%%%%%%%%%%%%%%%%%%%%%%%%%%%%%%%%%%%%%%%%%%%%%%%%%%%%%%%%
\begin{figure}[t]
\centering
\includegraphics[width=\columnwidth]{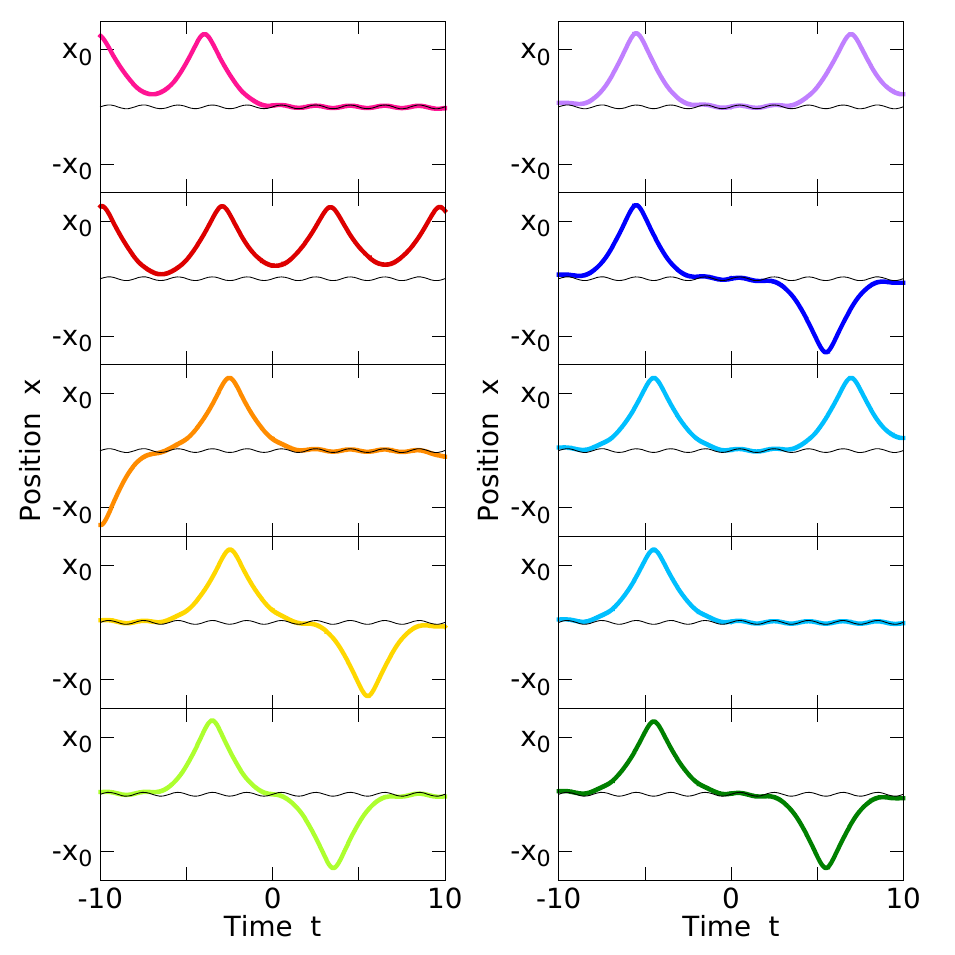}
\caption{%
Selection of ten typical trajectories in the potential~\eqref{eq:potential} 
whose initial phase space coordinates at $t=0$ correspond to minima of the LD 
(see \FIG~\ref{fig:minima}).
Each of the trajectories is characterized by a combination of small 
oscillations at the barrier (oscillations with amplitude to $x<1$) in addition 
to large 
oscillations in the wells (oscillations with amplitude $x>2$).
In each plot, the periodic TS trajectory is also shown as a black line for 
comparison.
}
\label{fig:trajectories}
\end{figure}
%%%%%%%%%%%%%%%%%%%%%%%%%%%%%%%%%%%%%%%%%%%%%%%%%%%%%%%%%%%%%%%%%%%%%%%%%%%%%%%

To obtain a precise understanding of the background and the occurrence of the 
observed LD structure as well as their values at the minima, we present in 
\FIG~\ref{fig:trajectories} a selection of ten typical trajectories which are 
obtained from LD minima.
(Note that it is here not important which of the trajectories belongs to which 
minimum, because the general behavior of the trajectories explains the LD 
structure and that is the same in all the cases.)
We see in \FIG~\ref{fig:trajectories} that the trajectories corresponding to 
minima of the LD are characterized by a combination of small 
oscillations at the barrier (oscillations with amplitude $x<1$) in addition 
to large oscillations in the wells (oscillations with amplitude $x>2$).
The periodic TS trajectory is also shown as a black line for comparison.
Each of these oscillations corresponds to a certain arg length of the 
trajectory and if we denote by $\LDbo$ the arc length of one oscillation at the 
barrier as well as by $\LDwo$ one oscillation in the well, the complete LD 
value corresponding to one of its minima is
\begin{equation}
 \LD \approx m \LDbo + n \LDwo
 \label{eq:LD-self-similar}
\end{equation}
with $m,n=0,1,2,\ldots$
The `approximate equal' sign is here intended to take into account that small 
deviations from the exact combinations occur in practice, because the single 
oscillations need to be connected smoothly according to the underlying 
dynamical equations.
Equation~\eqref{eq:LD-self-similar} directly explains the self-similar 
structure of the LD because any linear combination of the LD contributions 
$\LDbo$ and $\LDwo$ leads to a (local) minimum if those values themselves 
correspond to local minima.
In addition, this relation also explains the step structure presented in 
\FIG~\ref{fig:minima}(b) as a result of different integers $m,n$ while the 
small slope seen within some of the steps comes from fulfilling the dynamical 
boundary conditions between barrier and well oscillations.
In addition, \FIG~\ref{fig:trajectories} also makes clear that we have two 
important time scales in our system:
one of them corresponds to the period of the barrier top (fast, 
small-amplitude oscillations) and the other one is the recurrence time of a 
trajectory (slow, big-amplitude oscillations) which is roughly four times as 
large.

%%%%%%%%%%%%%%%%%%%%%%%%%%%%%%%%%%%%%%%%%%%%%%%%%%%%%%%%%%%%%%%%%%%%%%%%%%%%%%%
\begin{figure*}[t]
\includegraphics[width=\textwidth]{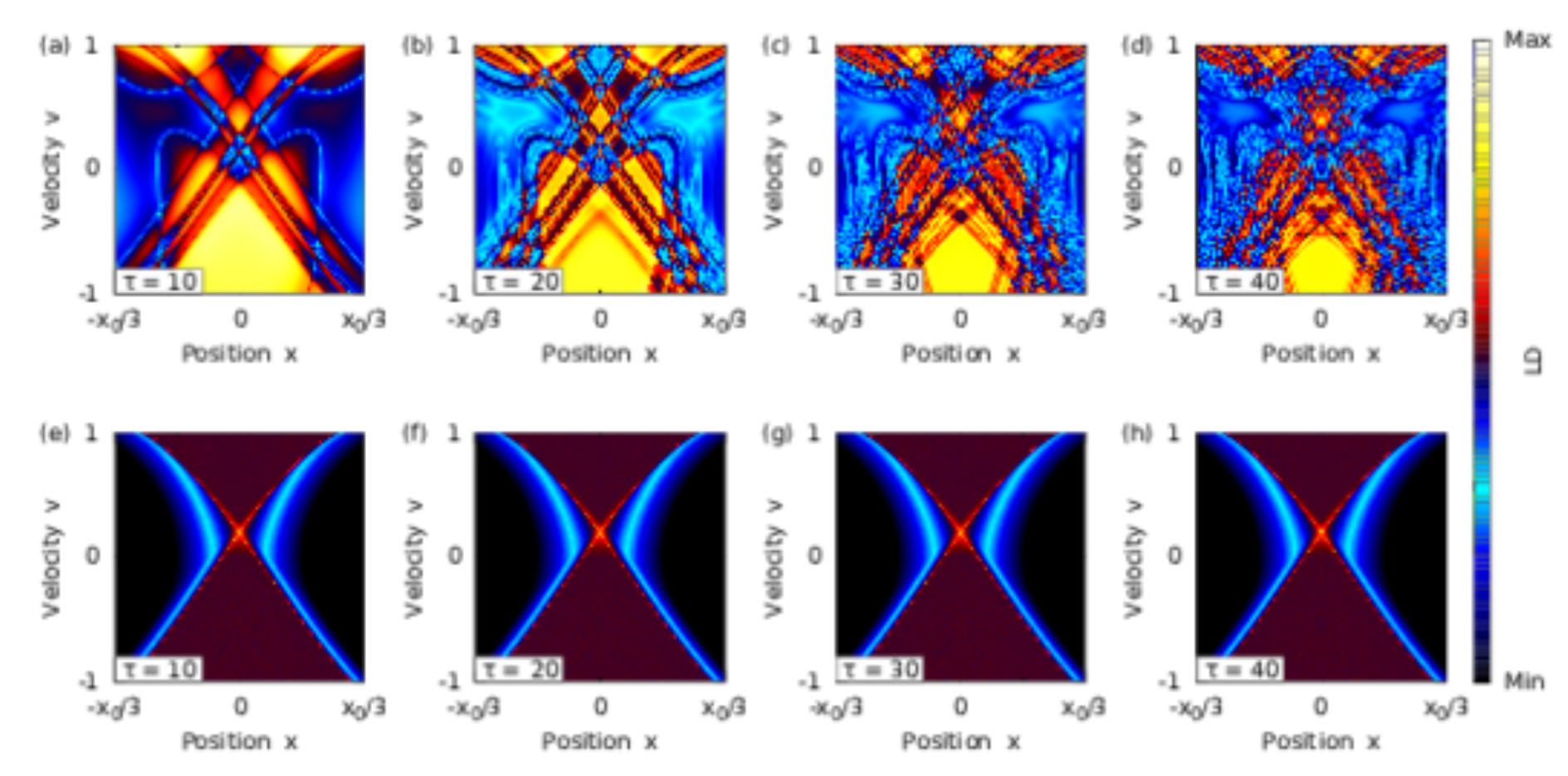}
\caption{%
Comparison of LD phase space portraits using the standard LD definition 
according to \EQ~\eqref{eq:LD} (top row) and the modified definition with 
variable integration time, \EQ~\eqref{eq:def-tau+-} (bottom row).
(The $x$-axes cover the range $-1 \leq x \leq 1$.)
}
\label{fig:LDs}
\end{figure*}
%%%%%%%%%%%%%%%%%%%%%%%%%%%%%%%%%%%%%%%%%%%%%%%%%%%%%%%%%%%%%%%%%%%%%%%%%%%%%%%

We have, so far, investigated and explained the self-similar LD structure close 
to the barrier top in the system with finite reactant and product wells, and we 
have seen that it is the properties of the underlying trajectories which result 
in these observations.
As we have already mentioned in the introduction, it is a major purpose of the 
LD method to provide a construction scheme for finding the TS trajectory.
With the results previously presented in this work, 
\ie~the occurrence of a huge 
number of local LD minima close to the barrier region (or formally an infinite 
number in the limit $\tau\to\infty$), one can easily imagine that this can 
result in significant problems in the application of the method.
Also, the result from \FIG~\ref{fig:minima}(b) that the desired local LD 
minimum is that with the smallest LD value, \ie~the global minimum in the 
barrier region, is only of minor help, because of the following reasons:
First, its systematic search then requires global optimization procedures, 
which are not easy to apply, especially in context with the observed 
self-similar minimum structure.
Second, only the comparison of the LD values from a selection of numerically 
obtained points is not sufficient, because there is no guarantee that the 
desired minima is part of the selection so that there cannot be another minimum 
with an even smaller LD value.

In the following, we show how the problem of additional (local) 
minima is solved using the modified LD definition according to 
\EQS~\eqref{eq:LD-time-interval-redef} and \eqref{eq:def-tau+-}.
A comparison between the corresponding LD phase space portraits is presented in 
\FIG~\ref{fig:LDs} using the standard LD definition in the top row and the 
modified definition with variable integration time in the bottom row.
As can be seen in the top row and as we have discussed above, a complicated 
structure of the stable and unstable manifolds or the LD valleys is present with 
the standard definition of the LD, \EQ~\eqref{eq:LD}.
This structure becomes more and more complicated with an increasing number of 
local LD minima if the integration time $\tau$ is increased (left to right for 
the values $\tau=10, 20, 30, 40$). 
The reason for this is that longer integration times allow for more reflections 
(and therefore global recrossings)
which, then, induce the details in the substructure.

The occurrence of the complicated LD structure is immediately suppressed if the 
variable integration times~\eqref{eq:def-tau+-} are used as shown in the bottom 
row.
Here, only single, minorly curved lines can be observed for the manifolds 
$\Wsu$.
Moreover, their intersection is clearly visible and not accompanied by 
additional, local minima.
This observation especially holds independently of the maximum integration time 
$\tau$, so that the LD surface does not get more complicated for increasing 
integration time.
Note that it is a consequence of the trajectory cut-off that the LD surface 
apart from the intersection of the manifolds is now characterized by very small 
LD values while they naturally get larger if one approaches the intersection.
It therefore appears from the color map of the figure that the desired point 
has turned into a maximum of the LD.
However, the relation \eqref{eq:LD-min} is still true because there is a very 
sharp minimum at the intersection of the manifolds (which has an extension 
smaller than the resolution of the figure).
We emphasize that the cut-off of the integration time as defined in 
\EQ~\eqref{eq:def-tau+-} does not have any effect on the position of the global 
LD minimum, \ie~the initial conditions of the TS trajectory, because this 
trajectory does not leave the barrier region so that the cut-off is never 
applied to the TS trajectory.

%%%%%%%%%%%%%%%%%%%%%%%%%%%%%%%%%%%%%%%%%%%%%%%%%%%%%%%%%%%%%%%%%%%%%%%%%%%%%%%
\begin{figure}[t]
\includegraphics[width=\columnwidth]{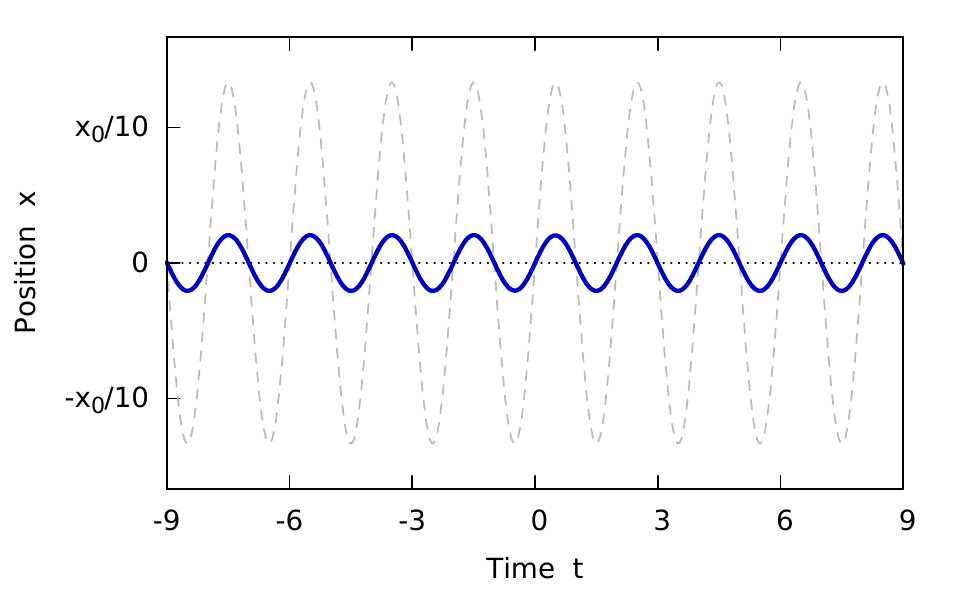}
\caption{%
Time evolution of the TS trajectory (solid blue line) and the barrier position 
$\xb$ (dashed gray line).
The initial conditions $(x=0, v=0.19738)$ of the TS trajectory at $t=0$ 
have been obtained from the minimum of the LD according to \EQ~\eqref{eq:LD-min}
}
\label{fig:TS-trajectory}
\end{figure}

%%%%%%%%%%%%%%%%%%%%%%%%%%%%%%%%%%%%%%%%%%%%%%%%%%%%%%%%%%%%%%%%%%%%%%%%%%%%%%%

The TS trajectory which corresponds to the \emph{single} minimum shown in the 
bottom row of \FIG~\ref{fig:LDs} is finally presented as the solid blue line 
% \comment{Matze: don't u mean blue line here?}
in \FIG~\ref{fig:TS-trajectory} (the dashed lines shows the time evolution of 
the barrier top).
As expected, the trajectory does not leave the barrier region but remains 
within the range of the barrier amplitude ($-0.4\leq x \leq 0.4$) which 
verifies the success of the procedure.

\section{Conclusion and outlook}

In this paper, we have investigated the phase space structure of a 
time-dependent double-well potential with special regard to the local minima of 
the LD and their connection to the TS trajectory.
We have seen that reflections of the particle at the potential walls lead 
to global recrossings and therefore to a 
self-similar LD structure with a huge number of local minima (or formally an 
infinite number in the limit $\tau\to\infty$).
We have demonstrated that this structure is directly related to the 
trajectories with
each local minimum corresponding 
to a linear combination of 
barrier and well oscillations.
This complicated structure naively
appears to be a significant obstacle in 
the application of the LD formalism to the
construction of TS trajectories.
It can be overcome, however,  
by a simple modification of the LD definition in a way that the 
underlying trajectories are cut off as soon as they leave the barrier region.
As a consequence, only local crossings of the barrier are taken into account.
This improved formalism is especially important to the application of the 
LD-DS method to systems 
in high-dimensional phase space
because the number of global recrossings---viz. oscillations---increases
with dimensionality.

The chaotic and self-similar behavior of this driven system 
echo those seen in atom-diatom reactions by
Tyapan and Jaffe.\cite{jaffe94,jaffe95}
As in their work, the DS on the TS trajectory is a homoclinic tangle that 
can be characterized as a fractal tiling.
Future work could and should obtain the scaling law for the tiling and
use it to obtain renormalized rates from the sum of fluxes on each tile.
Such an approach holds promise in obtaining the rate using the LD-DS scheme
locally while addressing increasingly higher-dimensional and more complex
barriers.

\EDITS{The LD-DS method presented 
in this paper can also be used to revisit the earlier
uses of the LD.
For example, in Ref.~\cite{Mancho2013} the fractal-like structures 
we found in our work
is already slightly visible as one cannot identify a single intersection 
of the manifolds near the saddle. 
The modification of the LD needed here to obtain
the TS trajectory and clearly reveal the fractal structure
should be applicable to their use of the LD for obtaining 
fixed dividing surfaces
when the latter LD is beset by the existencs of many local minima due
to recurrences.
}

\begin{acknowledgments}
AJ acknowledges the Alexander von Humboldt Foundation, Germany, 
for support through a Feodor Lynen Fellowship.
RH's contribution to this work was supported by
the National Science Foundation (NSF) through Grant
No.~CHE-1700749.
This collaboration has also benefited from support
by the people mobility programs, and most recently by the
European Union’s Horizon 2020 research and innovation
programm under Grant Agreement No.~734557.
We thank reviewer 2 for pointing out the approach by Tiyapan and
Jaff\'e\cite{jaffe94,jaffe95} to resolve the scaling laws of the fractal
tiling of the dividing surface as a possible future dirction
for resolving rates using the LD-DS method.
\end{acknowledgments}

\section*{References}
\bibliography{p99}

%merlin.mbs aipnum4-1.bst 2010-07-25 4.21a (PWD, AO, DPC) hacked
%Control: key (0)
%Control: author (8) initials jnrlst
%Control: editor formatted (1) identically to author
%Control: production of article title (-1) disabled
%Control: page (0) single
%Control: year (1) truncated
%Control: production of eprint (0) enabled
\begin{thebibliography}{50}%
\makeatletter
\providecommand \@ifxundefined [1]{%
 \@ifx{#1\undefined}
}%
\providecommand \@ifnum [1]{%
 \ifnum #1\expandafter \@firstoftwo
 \else \expandafter \@secondoftwo
 \fi
}%
\providecommand \@ifx [1]{%
 \ifx #1\expandafter \@firstoftwo
 \else \expandafter \@secondoftwo
 \fi
}%
\providecommand \natexlab [1]{#1}%
\providecommand \enquote  [1]{``#1''}%
\providecommand \bibnamefont  [1]{#1}%
\providecommand \bibfnamefont [1]{#1}%
\providecommand \citenamefont [1]{#1}%
\providecommand \href@noop [0]{\@secondoftwo}%
\providecommand \href [0]{\begingroup \@sanitize@url \@href}%
\providecommand \@href[1]{\@@startlink{#1}\@@href}%
\providecommand \@@href[1]{\endgroup#1\@@endlink}%
\providecommand \@sanitize@url [0]{\catcode `\\12\catcode `\$12\catcode
  `\&12\catcode `\#12\catcode `\^12\catcode `\_12\catcode `\%12\relax}%
\providecommand \@@startlink[1]{}%
\providecommand \@@endlink[0]{}%
\providecommand \url  [0]{\begingroup\@sanitize@url \@url }%
\providecommand \@url [1]{\endgroup\@href {#1}{\urlprefix }}%
\providecommand \urlprefix  [0]{URL }%
\providecommand \Eprint [0]{\href }%
\providecommand \doibase [0]{http://dx.doi.org/}%
\providecommand \selectlanguage [0]{\@gobble}%
\providecommand \bibinfo  [0]{\@secondoftwo}%
\providecommand \bibfield  [0]{\@secondoftwo}%
\providecommand \translation [1]{[#1]}%
\providecommand \BibitemOpen [0]{}%
\providecommand \bibitemStop [0]{}%
\providecommand \bibitemNoStop [0]{.\EOS\space}%
\providecommand \EOS [0]{\spacefactor3000\relax}%
\providecommand \BibitemShut  [1]{\csname bibitem#1\endcsname}%
\let\auto@bib@innerbib\@empty
%</preamble>
\bibitem [{\citenamefont {Yamanouchi}(2002)}]{Yamanouchi2002}%
  \BibitemOpen
  \bibfield  {author} {\bibinfo {author} {\bibfnamefont {K.}~\bibnamefont
  {Yamanouchi}},\ }\href@noop {} {\bibfield  {journal} {\bibinfo  {journal}
  {Science}\ }\textbf {\bibinfo {volume} {295}},\ \bibinfo {pages} {1659}
  (\bibinfo {year} {2002})}\BibitemShut {NoStop}%
\bibitem [{\citenamefont {Hershkovits}\ and\ \citenamefont
  {Hernandez}(2005)}]{hern05a}%
  \BibitemOpen
  \bibfield  {author} {\bibinfo {author} {\bibfnamefont {E.}~\bibnamefont
  {Hershkovits}}\ and\ \bibinfo {author} {\bibfnamefont {R.}~\bibnamefont
  {Hernandez}},\ }\href@noop {} {\bibfield  {journal} {\bibinfo  {journal} {J.
  Chem. Phys.}\ }\textbf {\bibinfo {volume} {122}},\ \bibinfo {pages} {014509}
  (\bibinfo {year} {2005})}\BibitemShut {NoStop}%
\bibitem [{\citenamefont {Sussman}\ \emph {et~al.}(2006)\citenamefont
  {Sussman}, \citenamefont {Townsend}, \citenamefont {Ivanov},\ and\
  \citenamefont {Stolow}}]{Sussman2006}%
  \BibitemOpen
  \bibfield  {author} {\bibinfo {author} {\bibfnamefont {B.~J.}\ \bibnamefont
  {Sussman}}, \bibinfo {author} {\bibfnamefont {D.}~\bibnamefont {Townsend}},
  \bibinfo {author} {\bibfnamefont {M.~Y.}\ \bibnamefont {Ivanov}}, \ and\
  \bibinfo {author} {\bibfnamefont {A.}~\bibnamefont {Stolow}},\ }\href@noop {}
  {\bibfield  {journal} {\bibinfo  {journal} {Science}\ }\textbf {\bibinfo
  {volume} {314}},\ \bibinfo {pages} {278} (\bibinfo {year}
  {2006})}\BibitemShut {NoStop}%
\bibitem [{\citenamefont {Kawai}\ \emph {et~al.}(2007)\citenamefont {Kawai},
  \citenamefont {Bandrauk}, \citenamefont {Jaff{\'e}}, \citenamefont {Bartsch},
  \citenamefont {Palaci{\'a}n},\ and\ \citenamefont {Uzer}}]{Kawai07}%
  \BibitemOpen
  \bibfield  {author} {\bibinfo {author} {\bibfnamefont {S.}~\bibnamefont
  {Kawai}}, \bibinfo {author} {\bibfnamefont {A.~D.}\ \bibnamefont {Bandrauk}},
  \bibinfo {author} {\bibfnamefont {C.}~\bibnamefont {Jaff{\'e}}}, \bibinfo
  {author} {\bibfnamefont {T.}~\bibnamefont {Bartsch}}, \bibinfo {author}
  {\bibfnamefont {J.}~\bibnamefont {Palaci{\'a}n}}, \ and\ \bibinfo {author}
  {\bibfnamefont {T.}~\bibnamefont {Uzer}},\ }\href@noop {} {\bibfield
  {journal} {\bibinfo  {journal} {J. Chem. Phys.}\ }\textbf {\bibinfo {volume}
  {126}},\ \bibinfo {pages} {164306} (\bibinfo {year} {2007})}\BibitemShut
  {NoStop}%
\bibitem [{\citenamefont {Kawai}\ and\ \citenamefont
  {Komatsuzaki}(2011)}]{Kawai11laser}%
  \BibitemOpen
  \bibfield  {author} {\bibinfo {author} {\bibfnamefont {S.}~\bibnamefont
  {Kawai}}\ and\ \bibinfo {author} {\bibfnamefont {T.}~\bibnamefont
  {Komatsuzaki}},\ }\href@noop {} {\bibfield  {journal} {\bibinfo  {journal}
  {J. Chem. Phys.}\ }\textbf {\bibinfo {volume} {134}},\ \bibinfo {pages}
  {024317} (\bibinfo {year} {2011})}\BibitemShut {NoStop}%
\bibitem [{\citenamefont {Sethi}\ and\ \citenamefont
  {Keshavamurthy}(2009)}]{Keshavamurthy2009}%
  \BibitemOpen
  \bibfield  {author} {\bibinfo {author} {\bibfnamefont {A.}~\bibnamefont
  {Sethi}}\ and\ \bibinfo {author} {\bibfnamefont {S.}~\bibnamefont
  {Keshavamurthy}},\ }\href@noop {} {\bibfield  {journal} {\bibinfo  {journal}
  {Phys. Rev. A}\ }\textbf {\bibinfo {volume} {79}},\ \bibinfo {pages} {033416}
  (\bibinfo {year} {2009})}\BibitemShut {NoStop}%
\bibitem [{\citenamefont {Patra}\ and\ \citenamefont
  {Keshavamurthy}(2015)}]{Keshavamurthy2015}%
  \BibitemOpen
  \bibfield  {author} {\bibinfo {author} {\bibfnamefont {S.}~\bibnamefont
  {Patra}}\ and\ \bibinfo {author} {\bibfnamefont {S.}~\bibnamefont
  {Keshavamurthy}},\ }\href@noop {} {\bibfield  {journal} {\bibinfo  {journal}
  {Chem. Phys. Lett.}\ }\textbf {\bibinfo {volume} {634}},\ \bibinfo {pages}
  {1} (\bibinfo {year} {2015})}\BibitemShut {NoStop}%
\bibitem [{\citenamefont {Revuelta}, \citenamefont {Chac\'on},\ and\
  \citenamefont {Borondo}(2015)}]{Revuelta2015}%
  \BibitemOpen
  \bibfield  {author} {\bibinfo {author} {\bibfnamefont {F.}~\bibnamefont
  {Revuelta}}, \bibinfo {author} {\bibfnamefont {R.}~\bibnamefont {Chac\'on}},
  \ and\ \bibinfo {author} {\bibfnamefont {F.}~\bibnamefont {Borondo}},\
  }\href@noop {} {\bibfield  {journal} {\bibinfo  {journal} {Europhys. Lett.}\
  }\textbf {\bibinfo {volume} {110}},\ \bibinfo {pages} {40007} (\bibinfo
  {year} {2015})}\BibitemShut {NoStop}%
\bibitem [{\citenamefont {Pitzer}, \citenamefont {Smith},\ and\ \citenamefont
  {Eyring}(1962)}]{pitzer}%
  \BibitemOpen
  \bibfield  {author} {\bibinfo {author} {\bibfnamefont {K.~S.}\ \bibnamefont
  {Pitzer}}, \bibinfo {author} {\bibfnamefont {F.~T.}\ \bibnamefont {Smith}}, \
  and\ \bibinfo {author} {\bibfnamefont {H.}~\bibnamefont {Eyring}},\
  }\href@noop {} {\emph {\bibinfo {title} {The Transition State}}},\ Special
  Publ.\ (\bibinfo  {publisher} {Chemical Society},\ \bibinfo {address}
  {London},\ \bibinfo {year} {1962})\ p.~\bibinfo {pages} {53}\BibitemShut
  {NoStop}%
\bibitem [{\citenamefont {Pechukas}(1981)}]{pechukas1981}%
  \BibitemOpen
  \bibfield  {author} {\bibinfo {author} {\bibfnamefont {P.}~\bibnamefont
  {Pechukas}},\ }\href@noop {} {\bibfield  {journal} {\bibinfo  {journal}
  {Annu. Rev. Phys. Chem.}\ }\textbf {\bibinfo {volume} {32}},\ \bibinfo
  {pages} {159} (\bibinfo {year} {1981})}\BibitemShut {NoStop}%
\bibitem [{\citenamefont {Garrett}\ and\ \citenamefont
  {Truhlar}(1979)}]{truh79}%
  \BibitemOpen
  \bibfield  {author} {\bibinfo {author} {\bibfnamefont {B.~C.}\ \bibnamefont
  {Garrett}}\ and\ \bibinfo {author} {\bibfnamefont {D.~G.}\ \bibnamefont
  {Truhlar}},\ }\href@noop {} {\bibfield  {journal} {\bibinfo  {journal} {J.
  Phys. Chem.}\ }\textbf {\bibinfo {volume} {83}},\ \bibinfo {pages} {1052}
  (\bibinfo {year} {1979})}\BibitemShut {NoStop}%
\bibitem [{\citenamefont {Truhlar}, \citenamefont {Issacson},\ and\
  \citenamefont {Garrett}(1985)}]{truh85}%
  \BibitemOpen
  \bibfield  {author} {\bibinfo {author} {\bibfnamefont {D.~G.}\ \bibnamefont
  {Truhlar}}, \bibinfo {author} {\bibfnamefont {A.~D.}\ \bibnamefont
  {Issacson}}, \ and\ \bibinfo {author} {\bibfnamefont {B.~C.}\ \bibnamefont
  {Garrett}},\ }\enquote {\bibinfo {title} {Theory of chemical reaction
  dynamics},}\ \ (\bibinfo  {publisher} {CRC Press},\ \bibinfo {address} {Boca
  Raton, FL},\ \bibinfo {year} {1985})\ pp.\ \bibinfo {pages}
  {65--137}\BibitemShut {NoStop}%
\bibitem [{\citenamefont {Natanson}\ \emph {et~al.}(1991)\citenamefont
  {Natanson}, \citenamefont {Garrett}, \citenamefont {Truong}, \citenamefont
  {Joseph},\ and\ \citenamefont {Truhlar}}]{truhlar91}%
  \BibitemOpen
  \bibfield  {author} {\bibinfo {author} {\bibfnamefont {G.~A.}\ \bibnamefont
  {Natanson}}, \bibinfo {author} {\bibfnamefont {B.~C.}\ \bibnamefont
  {Garrett}}, \bibinfo {author} {\bibfnamefont {T.~N.}\ \bibnamefont {Truong}},
  \bibinfo {author} {\bibfnamefont {T.}~\bibnamefont {Joseph}}, \ and\ \bibinfo
  {author} {\bibfnamefont {D.~G.}\ \bibnamefont {Truhlar}},\ }\href@noop {}
  {\bibfield  {journal} {\bibinfo  {journal} {J. Chem. Phys.}\ }\textbf
  {\bibinfo {volume} {94}},\ \bibinfo {pages} {7875} (\bibinfo {year}
  {1991})}\BibitemShut {NoStop}%
\bibitem [{\citenamefont {Truhlar}, \citenamefont {Garrett},\ and\
  \citenamefont {Klippenstein}(1996)}]{truh96}%
  \BibitemOpen
  \bibfield  {author} {\bibinfo {author} {\bibfnamefont {D.~G.}\ \bibnamefont
  {Truhlar}}, \bibinfo {author} {\bibfnamefont {B.~C.}\ \bibnamefont
  {Garrett}}, \ and\ \bibinfo {author} {\bibfnamefont {S.~J.}\ \bibnamefont
  {Klippenstein}},\ }\href@noop {} {\bibfield  {journal} {\bibinfo  {journal}
  {J. Phys. Chem.}\ }\textbf {\bibinfo {volume} {100}},\ \bibinfo {pages}
  {12771} (\bibinfo {year} {1996})}\BibitemShut {NoStop}%
\bibitem [{\citenamefont {Truhlar}\ and\ \citenamefont
  {Garrett}(2000)}]{truh2000}%
  \BibitemOpen
  \bibfield  {author} {\bibinfo {author} {\bibfnamefont {D.~G.}\ \bibnamefont
  {Truhlar}}\ and\ \bibinfo {author} {\bibfnamefont {B.~C.}\ \bibnamefont
  {Garrett}},\ }\href@noop {} {\bibfield  {journal} {\bibinfo  {journal} {J.
  Phys. Chem. B}\ }\textbf {\bibinfo {volume} {104}},\ \bibinfo {pages} {1069}
  (\bibinfo {year} {2000})}\BibitemShut {NoStop}%
\bibitem [{\citenamefont {Komatsuzaki}\ and\ \citenamefont
  {Berry}(2001)}]{Komatsuzaki2001}%
  \BibitemOpen
  \bibfield  {author} {\bibinfo {author} {\bibfnamefont {T.}~\bibnamefont
  {Komatsuzaki}}\ and\ \bibinfo {author} {\bibfnamefont {R.~S.}\ \bibnamefont
  {Berry}},\ }\href@noop {} {\bibfield  {journal} {\bibinfo  {journal} {Proc.
  Natl. Acad. Sci. U.S.A.}\ }\textbf {\bibinfo {volume} {98}},\ \bibinfo
  {pages} {7666} (\bibinfo {year} {2001})}\BibitemShut {NoStop}%
\bibitem [{\citenamefont {Waalkens}, \citenamefont {Schubert},\ and\
  \citenamefont {Wiggins}(2008)}]{Waalkens2008}%
  \BibitemOpen
  \bibfield  {author} {\bibinfo {author} {\bibfnamefont {H.}~\bibnamefont
  {Waalkens}}, \bibinfo {author} {\bibfnamefont {R.}~\bibnamefont {Schubert}},
  \ and\ \bibinfo {author} {\bibfnamefont {S.}~\bibnamefont {Wiggins}},\
  }\href@noop {} {\bibfield  {journal} {\bibinfo  {journal} {Nonlinearity}\
  }\textbf {\bibinfo {volume} {21}},\ \bibinfo {pages} {R1} (\bibinfo {year}
  {2008})}\BibitemShut {NoStop}%
\bibitem [{\citenamefont {Bartsch}\ \emph {et~al.}(2008)\citenamefont
  {Bartsch}, \citenamefont {Moix}, \citenamefont {Hernandez}, \citenamefont
  {Kawai},\ and\ \citenamefont {Uzer}}]{hern08d}%
  \BibitemOpen
  \bibfield  {author} {\bibinfo {author} {\bibfnamefont {T.}~\bibnamefont
  {Bartsch}}, \bibinfo {author} {\bibfnamefont {J.~M.}\ \bibnamefont {Moix}},
  \bibinfo {author} {\bibfnamefont {R.}~\bibnamefont {Hernandez}}, \bibinfo
  {author} {\bibfnamefont {S.}~\bibnamefont {Kawai}}, \ and\ \bibinfo {author}
  {\bibfnamefont {T.}~\bibnamefont {Uzer}},\ }\href@noop {} {\bibfield
  {journal} {\bibinfo  {journal} {Adv. Chem. Phys.}\ }\textbf {\bibinfo
  {volume} {140}},\ \bibinfo {pages} {191} (\bibinfo {year}
  {2008})}\BibitemShut {NoStop}%
\bibitem [{\citenamefont {Kawai}\ and\ \citenamefont
  {Komatsuzaki}(2010)}]{Komatsuzaki2010}%
  \BibitemOpen
  \bibfield  {author} {\bibinfo {author} {\bibfnamefont {S.}~\bibnamefont
  {Kawai}}\ and\ \bibinfo {author} {\bibfnamefont {T.}~\bibnamefont
  {Komatsuzaki}},\ }\href@noop {} {\bibfield  {journal} {\bibinfo  {journal}
  {Phys. Rev. Lett.}\ }\textbf {\bibinfo {volume} {105}},\ \bibinfo {pages}
  {048304} (\bibinfo {year} {2010})}\BibitemShut {NoStop}%
\bibitem [{\citenamefont {Hernandez}, \citenamefont {Bartsch},\ and\
  \citenamefont {Uzer}(2010)}]{hern10a}%
  \BibitemOpen
  \bibfield  {author} {\bibinfo {author} {\bibfnamefont {R.}~\bibnamefont
  {Hernandez}}, \bibinfo {author} {\bibfnamefont {T.}~\bibnamefont {Bartsch}},
  \ and\ \bibinfo {author} {\bibfnamefont {T.}~\bibnamefont {Uzer}},\
  }\href@noop {} {\bibfield  {journal} {\bibinfo  {journal} {Chem. Phys.}\
  }\textbf {\bibinfo {volume} {370}},\ \bibinfo {pages} {270} (\bibinfo {year}
  {2010})}\BibitemShut {NoStop}%
\bibitem [{\citenamefont {Sharia}\ and\ \citenamefont
  {Henkelman}(2016)}]{Henkelman2016}%
  \BibitemOpen
  \bibfield  {author} {\bibinfo {author} {\bibfnamefont {O.}~\bibnamefont
  {Sharia}}\ and\ \bibinfo {author} {\bibfnamefont {G.}~\bibnamefont
  {Henkelman}},\ }\href@noop {} {\bibfield  {journal} {\bibinfo  {journal} {New
  J. Phys.}\ }\textbf {\bibinfo {volume} {18}},\ \bibinfo {pages} {013023}
  (\bibinfo {year} {2016})}\BibitemShut {NoStop}%
\bibitem [{\citenamefont {Pollak}\ and\ \citenamefont
  {Pechukas}(1978)}]{pollak78}%
  \BibitemOpen
  \bibfield  {author} {\bibinfo {author} {\bibfnamefont {E.}~\bibnamefont
  {Pollak}}\ and\ \bibinfo {author} {\bibfnamefont {P.}~\bibnamefont
  {Pechukas}},\ }\href@noop {} {\bibfield  {journal} {\bibinfo  {journal} {J.
  Chem. Phys.}\ }\textbf {\bibinfo {volume} {69}},\ \bibinfo {pages} {1218}
  (\bibinfo {year} {1978})}\BibitemShut {NoStop}%
\bibitem [{\citenamefont {Pechukas}\ and\ \citenamefont
  {Pollak}(1979)}]{pech79a}%
  \BibitemOpen
  \bibfield  {author} {\bibinfo {author} {\bibfnamefont {P.}~\bibnamefont
  {Pechukas}}\ and\ \bibinfo {author} {\bibfnamefont {E.}~\bibnamefont
  {Pollak}},\ }\href@noop {} {\bibfield  {journal} {\bibinfo  {journal} {J.
  Chem. Phys.}\ }\textbf {\bibinfo {volume} {71}},\ \bibinfo {pages} {2062}
  (\bibinfo {year} {1979})}\BibitemShut {NoStop}%
\bibitem [{\citenamefont {Hernandez}\ and\ \citenamefont
  {Miller}(1993)}]{hern93b}%
  \BibitemOpen
  \bibfield  {author} {\bibinfo {author} {\bibfnamefont {R.}~\bibnamefont
  {Hernandez}}\ and\ \bibinfo {author} {\bibfnamefont {W.~H.}\ \bibnamefont
  {Miller}},\ }\href@noop {} {\bibfield  {journal} {\bibinfo  {journal} {Chem.
  Phys. Lett.}\ }\textbf {\bibinfo {volume} {214}},\ \bibinfo {pages} {129}
  (\bibinfo {year} {1993})}\BibitemShut {NoStop}%
\bibitem [{\citenamefont {Hernandez}(1994)}]{hern94}%
  \BibitemOpen
  \bibfield  {author} {\bibinfo {author} {\bibfnamefont {R.}~\bibnamefont
  {Hernandez}},\ }\href@noop {} {\bibfield  {journal} {\bibinfo  {journal} {J.
  Chem. Phys.}\ }\textbf {\bibinfo {volume} {101}},\ \bibinfo {pages} {9534}
  (\bibinfo {year} {1994})}\BibitemShut {NoStop}%
\bibitem [{\citenamefont {Jaff{\'e}}, \citenamefont {Farrelly},\ and\
  \citenamefont {Uzer}(2000)}]{Jaffe00}%
  \BibitemOpen
  \bibfield  {author} {\bibinfo {author} {\bibfnamefont {C.}~\bibnamefont
  {Jaff{\'e}}}, \bibinfo {author} {\bibfnamefont {D.}~\bibnamefont {Farrelly}},
  \ and\ \bibinfo {author} {\bibfnamefont {T.}~\bibnamefont {Uzer}},\
  }\href@noop {} {\bibfield  {journal} {\bibinfo  {journal} {Phys. Rev. Lett.}\
  }\textbf {\bibinfo {volume} {84}},\ \bibinfo {pages} {610} (\bibinfo {year}
  {2000})}\BibitemShut {NoStop}%
\bibitem [{\citenamefont {Koon}\ \emph {et~al.}(2000)\citenamefont {Koon},
  \citenamefont {Lo}, \citenamefont {Marsden},\ and\ \citenamefont
  {Ross}}]{Koon00}%
  \BibitemOpen
  \bibfield  {author} {\bibinfo {author} {\bibfnamefont {W.~S.}\ \bibnamefont
  {Koon}}, \bibinfo {author} {\bibfnamefont {M.~W.}\ \bibnamefont {Lo}},
  \bibinfo {author} {\bibfnamefont {J.~E.}\ \bibnamefont {Marsden}}, \ and\
  \bibinfo {author} {\bibfnamefont {S.~D.}\ \bibnamefont {Ross}},\ }\href@noop
  {} {\bibfield  {journal} {\bibinfo  {journal} {Chaos}\ }\textbf {\bibinfo
  {volume} {10}},\ \bibinfo {pages} {427} (\bibinfo {year} {2000})}\BibitemShut
  {NoStop}%
\bibitem [{\citenamefont {Jaff{\'e}}\ \emph {et~al.}(2002)\citenamefont
  {Jaff{\'e}}, \citenamefont {Ross}, \citenamefont {Lo}, \citenamefont
  {Marsden}, \citenamefont {Farrelly},\ and\ \citenamefont {Uzer}}]{Jaffe02}%
  \BibitemOpen
  \bibfield  {author} {\bibinfo {author} {\bibfnamefont {C.}~\bibnamefont
  {Jaff{\'e}}}, \bibinfo {author} {\bibfnamefont {S.~D.}\ \bibnamefont {Ross}},
  \bibinfo {author} {\bibfnamefont {M.~W.}\ \bibnamefont {Lo}}, \bibinfo
  {author} {\bibfnamefont {J.}~\bibnamefont {Marsden}}, \bibinfo {author}
  {\bibfnamefont {D.}~\bibnamefont {Farrelly}}, \ and\ \bibinfo {author}
  {\bibfnamefont {T.}~\bibnamefont {Uzer}},\ }\href {\doibase
  10.1103/PhysRevLett.89.011101} {\bibfield  {journal} {\bibinfo  {journal}
  {Phys. Rev. Lett.}\ }\textbf {\bibinfo {volume} {89}},\ \bibinfo {pages}
  {011101} (\bibinfo {year} {2002})}\BibitemShut {NoStop}%
\bibitem [{\citenamefont {Uzer}\ \emph {et~al.}(2002)\citenamefont {Uzer},
  \citenamefont {Jaff{\'e}}, \citenamefont {Palaci{\'a}n}, \citenamefont
  {Yanguas},\ and\ \citenamefont {Wiggins}}]{Uzer02}%
  \BibitemOpen
  \bibfield  {author} {\bibinfo {author} {\bibfnamefont {T.}~\bibnamefont
  {Uzer}}, \bibinfo {author} {\bibfnamefont {C.}~\bibnamefont {Jaff{\'e}}},
  \bibinfo {author} {\bibfnamefont {J.}~\bibnamefont {Palaci{\'a}n}}, \bibinfo
  {author} {\bibfnamefont {P.}~\bibnamefont {Yanguas}}, \ and\ \bibinfo
  {author} {\bibfnamefont {S.}~\bibnamefont {Wiggins}},\ }\href@noop {}
  {\bibfield  {journal} {\bibinfo  {journal} {Nonlinearity}\ }\textbf {\bibinfo
  {volume} {15}},\ \bibinfo {pages} {957} (\bibinfo {year} {2002})}\BibitemShut
  {NoStop}%
\bibitem [{\citenamefont {Waalkens}\ and\ \citenamefont
  {Wiggins}(2004)}]{Waalkens04b}%
  \BibitemOpen
  \bibfield  {author} {\bibinfo {author} {\bibfnamefont {H.}~\bibnamefont
  {Waalkens}}\ and\ \bibinfo {author} {\bibfnamefont {S.}~\bibnamefont
  {Wiggins}},\ }\href@noop {} {\bibfield  {journal} {\bibinfo  {journal} {J.
  Phys. A}\ }\textbf {\bibinfo {volume} {37}},\ \bibinfo {pages} {L435}
  (\bibinfo {year} {2004})}\BibitemShut {NoStop}%
\bibitem [{\citenamefont {Jaff{\'e}}\ \emph {et~al.}(2005)\citenamefont
  {Jaff{\'e}}, \citenamefont {Kawai}, \citenamefont {Palaci{\'a}n},
  \citenamefont {Yanguas},\ and\ \citenamefont {Uzer}}]{Jaffe05}%
  \BibitemOpen
  \bibfield  {author} {\bibinfo {author} {\bibfnamefont {C.}~\bibnamefont
  {Jaff{\'e}}}, \bibinfo {author} {\bibfnamefont {S.}~\bibnamefont {Kawai}},
  \bibinfo {author} {\bibfnamefont {J.}~\bibnamefont {Palaci{\'a}n}}, \bibinfo
  {author} {\bibfnamefont {P.}~\bibnamefont {Yanguas}}, \ and\ \bibinfo
  {author} {\bibfnamefont {T.}~\bibnamefont {Uzer}},\ }\href@noop {} {\bibfield
   {journal} {\bibinfo  {journal} {Adv. Chem. Phys.}\ }\textbf {\bibinfo
  {volume} {130A}},\ \bibinfo {pages} {171} (\bibinfo {year}
  {2005})}\BibitemShut {NoStop}%
\bibitem [{\citenamefont {Li}\ \emph {et~al.}(2006)\citenamefont {Li},
  \citenamefont {Shoujiguchi}, \citenamefont {Toda},\ and\ \citenamefont
  {Komatsuzaki}}]{Li06prl}%
  \BibitemOpen
  \bibfield  {author} {\bibinfo {author} {\bibfnamefont {C.-B.}\ \bibnamefont
  {Li}}, \bibinfo {author} {\bibfnamefont {A.}~\bibnamefont {Shoujiguchi}},
  \bibinfo {author} {\bibfnamefont {M.}~\bibnamefont {Toda}}, \ and\ \bibinfo
  {author} {\bibfnamefont {T.}~\bibnamefont {Komatsuzaki}},\ }\href@noop {}
  {\bibfield  {journal} {\bibinfo  {journal} {Phys. Rev. Lett.}\ }\textbf
  {\bibinfo {volume} {97}},\ \bibinfo {pages} {028302(1)} (\bibinfo {year}
  {2006})}\BibitemShut {NoStop}%
\bibitem [{\citenamefont {Teramoto}, \citenamefont {Toda},\ and\ \citenamefont
  {Komatsuzaki}(2011)}]{Teramoto11}%
  \BibitemOpen
  \bibfield  {author} {\bibinfo {author} {\bibfnamefont {H.}~\bibnamefont
  {Teramoto}}, \bibinfo {author} {\bibfnamefont {M.}~\bibnamefont {Toda}}, \
  and\ \bibinfo {author} {\bibfnamefont {T.}~\bibnamefont {Komatsuzaki}},\
  }\href@noop {} {\bibfield  {journal} {\bibinfo  {journal} {Phys. Rev. Lett.}\
  }\textbf {\bibinfo {volume} {106}},\ \bibinfo {pages} {054101(1)} (\bibinfo
  {year} {2011})}\BibitemShut {NoStop}%
\bibitem [{\citenamefont {\ifmmode \mbox{\c{C}}\else
  \c{C}\fi{}ift\ifmmode~\mbox{\c{c}}\else \c{c}\fi{}i}\ and\ \citenamefont
  {Waalkens}(2013)}]{Waalkens13}%
  \BibitemOpen
  \bibfield  {author} {\bibinfo {author} {\bibfnamefont {U.}~\bibnamefont
  {\ifmmode \mbox{\c{C}}\else \c{C}\fi{}ift\ifmmode~\mbox{\c{c}}\else
  \c{c}\fi{}i}}\ and\ \bibinfo {author} {\bibfnamefont {H.}~\bibnamefont
  {Waalkens}},\ }\href@noop {} {\bibfield  {journal} {\bibinfo  {journal}
  {Phys. Rev. Lett.}\ }\textbf {\bibinfo {volume} {110}},\ \bibinfo {pages}
  {233201(1)} (\bibinfo {year} {2013})}\BibitemShut {NoStop}%
\bibitem [{\citenamefont {Bartsch}, \citenamefont {Uzer},\ and\ \citenamefont
  {Hernandez}(2005)}]{dawn05b}%
  \BibitemOpen
  \bibfield  {author} {\bibinfo {author} {\bibfnamefont {T.}~\bibnamefont
  {Bartsch}}, \bibinfo {author} {\bibfnamefont {T.}~\bibnamefont {Uzer}}, \
  and\ \bibinfo {author} {\bibfnamefont {R.}~\bibnamefont {Hernandez}},\
  }\href@noop {} {\bibfield  {journal} {\bibinfo  {journal} {J. Chem. Phys.}\
  }\textbf {\bibinfo {volume} {123}},\ \bibinfo {pages} {204102(1)} (\bibinfo
  {year} {2005})}\BibitemShut {NoStop}%
\bibitem [{\citenamefont {Bartsch}, \citenamefont {Hernandez},\ and\
  \citenamefont {Uzer}(2005)}]{dawn05a}%
  \BibitemOpen
  \bibfield  {author} {\bibinfo {author} {\bibfnamefont {T.}~\bibnamefont
  {Bartsch}}, \bibinfo {author} {\bibfnamefont {R.}~\bibnamefont {Hernandez}},
  \ and\ \bibinfo {author} {\bibfnamefont {T.}~\bibnamefont {Uzer}},\
  }\href@noop {} {\bibfield  {journal} {\bibinfo  {journal} {Phys. Rev. Lett.}\
  }\textbf {\bibinfo {volume} {95}},\ \bibinfo {pages} {058301(1)} (\bibinfo
  {year} {2005})}\BibitemShut {NoStop}%
\bibitem [{\citenamefont {Bartsch}\ \emph {et~al.}(2006)\citenamefont
  {Bartsch}, \citenamefont {Uzer}, \citenamefont {Moix},\ and\ \citenamefont
  {Hernandez}}]{hern06d}%
  \BibitemOpen
  \bibfield  {author} {\bibinfo {author} {\bibfnamefont {T.}~\bibnamefont
  {Bartsch}}, \bibinfo {author} {\bibfnamefont {T.}~\bibnamefont {Uzer}},
  \bibinfo {author} {\bibfnamefont {J.~M.}\ \bibnamefont {Moix}}, \ and\
  \bibinfo {author} {\bibfnamefont {R.}~\bibnamefont {Hernandez}},\ }\href@noop
  {} {\bibfield  {journal} {\bibinfo  {journal} {J. Chem. Phys.}\ }\textbf
  {\bibinfo {volume} {124}},\ \bibinfo {pages} {244310(01)} (\bibinfo {year}
  {2006})}\BibitemShut {NoStop}%
\bibitem [{\citenamefont {Craven}, \citenamefont {Bartsch},\ and\ \citenamefont
  {Hernandez}(2014{\natexlab{a}})}]{hern14b}%
  \BibitemOpen
  \bibfield  {author} {\bibinfo {author} {\bibfnamefont {G.~T.}\ \bibnamefont
  {Craven}}, \bibinfo {author} {\bibfnamefont {T.}~\bibnamefont {Bartsch}}, \
  and\ \bibinfo {author} {\bibfnamefont {R.}~\bibnamefont {Hernandez}},\
  }\href@noop {} {\bibfield  {journal} {\bibinfo  {journal} {Phys. Rev. E}\
  }\textbf {\bibinfo {volume} {89}},\ \bibinfo {pages} {040801(1)} (\bibinfo
  {year} {2014}{\natexlab{a}})}\BibitemShut {NoStop}%
\bibitem [{\citenamefont {Craven}, \citenamefont {Bartsch},\ and\ \citenamefont
  {Hernandez}(2014{\natexlab{b}})}]{hern14f}%
  \BibitemOpen
  \bibfield  {author} {\bibinfo {author} {\bibfnamefont {G.~T.}\ \bibnamefont
  {Craven}}, \bibinfo {author} {\bibfnamefont {T.}~\bibnamefont {Bartsch}}, \
  and\ \bibinfo {author} {\bibfnamefont {R.}~\bibnamefont {Hernandez}},\
  }\href@noop {} {\bibfield  {journal} {\bibinfo  {journal} {J. Chem. Phys.}\
  }\textbf {\bibinfo {volume} {141}},\ \bibinfo {pages} {041106(1)} (\bibinfo
  {year} {2014}{\natexlab{b}})}\BibitemShut {NoStop}%
\bibitem [{\citenamefont {Craven}, \citenamefont {Bartsch},\ and\ \citenamefont
  {Hernandez}(2015)}]{hern15a}%
  \BibitemOpen
  \bibfield  {author} {\bibinfo {author} {\bibfnamefont {G.~T.}\ \bibnamefont
  {Craven}}, \bibinfo {author} {\bibfnamefont {T.}~\bibnamefont {Bartsch}}, \
  and\ \bibinfo {author} {\bibfnamefont {R.}~\bibnamefont {Hernandez}},\
  }\href@noop {} {\bibfield  {journal} {\bibinfo  {journal} {J. Chem. Phys.}\
  }\textbf {\bibinfo {volume} {142}},\ \bibinfo {pages} {1} (\bibinfo {year}
  {2015})}\BibitemShut {NoStop}%
\bibitem [{\citenamefont {Kawai}\ and\ \citenamefont
  {Komatsuzaki}(2009)}]{Kawai2009a}%
  \BibitemOpen
  \bibfield  {author} {\bibinfo {author} {\bibfnamefont {S.}~\bibnamefont
  {Kawai}}\ and\ \bibinfo {author} {\bibfnamefont {T.}~\bibnamefont
  {Komatsuzaki}},\ }\href@noop {} {\bibfield  {journal} {\bibinfo  {journal}
  {J. Chem. Phys.}\ }\textbf {\bibinfo {volume} {131}},\ \bibinfo {pages}
  {224505(1)} (\bibinfo {year} {2009})}\BibitemShut {NoStop}%
\bibitem [{\citenamefont {Junginger}\ and\ \citenamefont
  {Hernandez}(2016)}]{hern16a}%
  \BibitemOpen
  \bibfield  {author} {\bibinfo {author} {\bibfnamefont {A.}~\bibnamefont
  {Junginger}}\ and\ \bibinfo {author} {\bibfnamefont {R.}~\bibnamefont
  {Hernandez}},\ }\href@noop {} {\bibfield  {journal} {\bibinfo  {journal} {J.
  Phys. Chem. B}\ }\textbf {\bibinfo {volume} {120}},\ \bibinfo {pages} {1720}
  (\bibinfo {year} {2016})}\BibitemShut {NoStop}%
\bibitem [{\citenamefont {Mendoza}\ and\ \citenamefont
  {Mancho}(2010)}]{Mancho2010}%
  \BibitemOpen
  \bibfield  {author} {\bibinfo {author} {\bibfnamefont {C.}~\bibnamefont
  {Mendoza}}\ and\ \bibinfo {author} {\bibfnamefont {A.~M.}\ \bibnamefont
  {Mancho}},\ }\href@noop {} {\bibfield  {journal} {\bibinfo  {journal} {Phys.
  Rev. Lett.}\ }\textbf {\bibinfo {volume} {105}},\ \bibinfo {pages} {038501}
  (\bibinfo {year} {2010})}\BibitemShut {NoStop}%
\bibitem [{\citenamefont {Mancho}\ \emph {et~al.}(2013)\citenamefont {Mancho},
  \citenamefont {Wiggins}, \citenamefont {Curbelo},\ and\ \citenamefont
  {Mendoza}}]{Mancho2013}%
  \BibitemOpen
  \bibfield  {author} {\bibinfo {author} {\bibfnamefont {A.~M.}\ \bibnamefont
  {Mancho}}, \bibinfo {author} {\bibfnamefont {S.}~\bibnamefont {Wiggins}},
  \bibinfo {author} {\bibfnamefont {J.}~\bibnamefont {Curbelo}}, \ and\
  \bibinfo {author} {\bibfnamefont {C.}~\bibnamefont {Mendoza}},\ }\href@noop
  {} {\bibfield  {journal} {\bibinfo  {journal} {Commun. Nonlinear Sci. Numer.
  Simul.}\ }\textbf {\bibinfo {volume} {18}},\ \bibinfo {pages} {3530 }
  (\bibinfo {year} {2013})}\BibitemShut {NoStop}%
\bibitem [{\citenamefont {Craven}\ and\ \citenamefont
  {Hernandez}(2015)}]{hern15e}%
  \BibitemOpen
  \bibfield  {author} {\bibinfo {author} {\bibfnamefont {G.~T.}\ \bibnamefont
  {Craven}}\ and\ \bibinfo {author} {\bibfnamefont {R.}~\bibnamefont
  {Hernandez}},\ }\href@noop {} {\bibfield  {journal} {\bibinfo  {journal}
  {Phys. Rev. Lett.}\ }\textbf {\bibinfo {volume} {115}},\ \bibinfo {pages}
  {148301} (\bibinfo {year} {2015})}\BibitemShut {NoStop}%
\bibitem [{\citenamefont {Craven}\ and\ \citenamefont
  {Hernandez}(2016)}]{hern16d}%
  \BibitemOpen
  \bibfield  {author} {\bibinfo {author} {\bibfnamefont {G.~T.}\ \bibnamefont
  {Craven}}\ and\ \bibinfo {author} {\bibfnamefont {R.}~\bibnamefont
  {Hernandez}},\ }\href@noop {} {\bibfield  {journal} {\bibinfo  {journal}
  {Phys. Chem. Chem. Phys.}\ }\textbf {\bibinfo {volume} {18}},\ \bibinfo
  {pages} {4008} (\bibinfo {year} {2016})}\BibitemShut {NoStop}%
\bibitem [{\citenamefont {Junginger}\ \emph {et~al.}(2016)\citenamefont
  {Junginger}, \citenamefont {Craven}, \citenamefont {Bartsch}, \citenamefont
  {Revuelta}, \citenamefont {Borondo}, \citenamefont {Benito},\ and\
  \citenamefont {Hernandez}}]{hern16h}%
  \BibitemOpen
  \bibfield  {author} {\bibinfo {author} {\bibfnamefont {A.}~\bibnamefont
  {Junginger}}, \bibinfo {author} {\bibfnamefont {G.~T.}\ \bibnamefont
  {Craven}}, \bibinfo {author} {\bibfnamefont {T.}~\bibnamefont {Bartsch}},
  \bibinfo {author} {\bibfnamefont {F.}~\bibnamefont {Revuelta}}, \bibinfo
  {author} {\bibfnamefont {F.}~\bibnamefont {Borondo}}, \bibinfo {author}
  {\bibfnamefont {R.~M.}\ \bibnamefont {Benito}}, \ and\ \bibinfo {author}
  {\bibfnamefont {R.}~\bibnamefont {Hernandez}},\ }\href@noop {} {\bibfield
  {journal} {\bibinfo  {journal} {Phys. Chem. Chem. Phys.}\ }\textbf {\bibinfo
  {volume} {18}},\ \bibinfo {pages} {30270} (\bibinfo {year}
  {2016})}\BibitemShut {NoStop}%
\bibitem [{\citenamefont {Mesele}\ and\ \citenamefont
  {Thompsona}(2016)}]{thompson16}%
  \BibitemOpen
  \bibfield  {author} {\bibinfo {author} {\bibfnamefont {O.~O.}\ \bibnamefont
  {Mesele}}\ and\ \bibinfo {author} {\bibfnamefont {W.~H.}\ \bibnamefont
  {Thompsona}},\ }\href@noop {} {\bibfield  {journal} {\bibinfo  {journal} {J.
  Chem. Phys.}\ }\textbf {\bibinfo {volume} {145}} (\bibinfo {year}
  {2016})}\BibitemShut {NoStop}%
\bibitem [{\citenamefont {Tiyapan}\ and\ \citenamefont
  {Jaff{\'e}}(1994)}]{jaffe94}%
  \BibitemOpen
  \bibfield  {author} {\bibinfo {author} {\bibfnamefont {A.}~\bibnamefont
  {Tiyapan}}\ and\ \bibinfo {author} {\bibfnamefont {C.}~\bibnamefont
  {Jaff{\'e}}},\ }\href@noop {} {\bibfield  {journal} {\bibinfo  {journal} {J.
  Chem. Phys.}\ }\textbf {\bibinfo {volume} {101}},\ \bibinfo {pages} {10393}
  (\bibinfo {year} {1994})}\BibitemShut {NoStop}%
\bibitem [{\citenamefont {Tiyapan}\ and\ \citenamefont
  {Jaff{\'e}}(1995)}]{jaffe95}%
  \BibitemOpen
  \bibfield  {author} {\bibinfo {author} {\bibfnamefont {A.}~\bibnamefont
  {Tiyapan}}\ and\ \bibinfo {author} {\bibfnamefont {C.}~\bibnamefont
  {Jaff{\'e}}},\ }\href@noop {} {\bibfield  {journal} {\bibinfo  {journal} {J.
  Chem. Phys.}\ }\textbf {\bibinfo {volume} {103}},\ \bibinfo {pages} {5499}
  (\bibinfo {year} {1995})}\BibitemShut {NoStop}%
\end{thebibliography}%
\end{document}